\documentclass[printer]{aa}
	
	\usepackage{footnote}
	\makesavenoteenv{tabular}
	\makesavenoteenv{table}

	\usepackage{booktabs}

	\usepackage{afterpage}
	\usepackage{float}
	\usepackage[flushleft]{threeparttable}
	\usepackage{tikz,pgfplots}
	
	\usetikzlibrary{external}
	\usepackage[capitalize]{cleveref}

	\usepackage{longtable}
	
	\usepackage[arrows=pgf]{mhchem}
	
	\pgfplotsset{compat=newest}
	
	\usepackage{csvsimple}
	\usepackage{lscape}
	
	\RequirePackage{txfonts}
	\RequirePackage{microtype}
	\RequirePackage{graphicx}
	\RequirePackage{booktabs}
	\RequirePackage{amsmath,amssymb}
	\RequirePackage{cleveref}
	\RequirePackage{pdfpages}
	\RequirePackage{textgreek}

	\newcommand*{\rangeto}{\textnormal{\hskip.1em\relax--\hskip.1em\relax}}
	\RequirePackage[separate-uncertainty=true,range-units=single,range-phrase=\rangeto,forbid-literal-units=true]{siunitx}

	\let\sun\odot
	
	\newcommand*{\solarmass}{\si{\solarmass}}
	\newcommand*{\solarluminosity}{\si{\solarluminosity}}
	\DeclareSIUnit\lightspeed{$c$}
	\DeclareSIUnit\rydberg{Ry}
	\DeclareSIUnit\erg{erg}
	\DeclareSIUnit\magnitude{mag}
	\DeclareSIUnit\jansky{Jy}
	\DeclareSIUnit\gauss{G}
	\DeclareSIUnit\h{$h$}
	\DeclareSIUnit\hseven{$h$_7}
	\DeclareSIUnit\parsec{pc}
	\DeclareSIUnit\au{AU}
	\DeclareSIUnit\year{yr}
	\DeclareSIUnit\solarluminosity{\ensuremath{L_\sun}}
	\DeclareSIUnit\solarmass{\ensuremath{M_\sun}}
	\DeclareSIUnit\jupitermass{\ensuremath{M_\mathrm{J}}}
	\DeclareSIUnit\solarmassinenergy{\ensuremath{M_\sun|c^2}}
	\DeclareSIUnit\solarradius{\ensuremath{R_\sun}}

	\DeclareSIUnit\arcsecond{as}
	\DeclareSIUnit\clight{\ensuremath c}
	\DeclareRobustCommand\SIfrac{\@ifstar\astrosym@SIfracplain\astrosym@SIfracparen}

	\newcommand{\myr}{\ensuremath{\mathrm{Myr}}}
	
	\newcommand{\cm}{\ensuremath{\mathrm{cm}}}
	\newcommand{\erg}{\ensuremath{\mathrm{erg}}}
	\newcommand{\persec}{\ensuremath{\mathrm{s}^{-1}}}
	
	\newcommand{\mjy}{\ensuremath{\mathrm{mJy}}}

	\newcommand{\msun}{\ensuremath{\mathrm{M}_\odot}}
	\newcommand{\rsun}{\ensuremath{\mathrm{R}_\odot}}

	\newcommand{\lsun}{\ensuremath{\mathrm{L}_\odot}}
	\newcommand{\mstar}{\ensuremath{M_*}}
	\newcommand{\lstar}{\ensuremath{L_*}}
	\newcommand{\rstar}{\ensuremath{R_*}}
	\newcommand{\mdisk}{\ensuremath{M_\mathrm{disk}}}

	\newcommand{\au}{\ensuremath{\mathrm{AU}}}
	\newcommand{\av}{\ensuremath{\mathrm{A}_V}}
	\newcommand{\micron}{\ensuremath{\mu\mathrm{m}}}
	
	\newcommand{\kelvin}{\ensuremath{\mathrm{K}}}
	\newcommand{\fuv}{\ensuremath{f_\mathrm{UV}}}
	
	\newcommand{\puv}{\ensuremath{p_\mathrm{UV}}}
	
	\newcommand{\lxr}{\ensuremath{L_\mathrm{X-ray}}}
	\newcommand{\rtaper}{\ensuremath{R_\mathrm{taper}}}
	\newcommand{\rin}{\ensuremath{R_\mathrm{in}}}
	\newcommand{\rout}{\ensuremath{R_\mathrm{out}}}
	\newcommand{\amin}{\ensuremath{a_\mathrm{min}}}
	\newcommand{\amax}{\ensuremath{a_\mathrm{max}}}

	\newcommand{\tgas}{\ensuremath{T_\mathrm{gas}}}

	\newcommand{\jwst}{\mbox{JWST}}
	\newcommand{\spitzer}{\mbox{Spitzer}}
	\newcommand{\flits}{\texttt{FLiTs}}
	\newcommand{\prodimo}{\texttt{ProDiMo}}

	\newcommand{\eelt}{\mbox{E-ELT}}

	\newcommand{\alma}{ALMA}
	\newcommand{\twopoppy}{\texttt{two-pop-py}}
	\newcommand{\mcmax}{\texttt{MCMax}}

	\newcommand{\ngrains}{\ensuremath{N_\mathrm{grains}}}
	\newcommand{\nrad}{\ensuremath{N_\mathrm{rad}}}
	\newcommand{\ntheta}{\ensuremath{N_\theta}}
	
	\newcommand{\tmax}{\ensuremath{t_\mathrm{max}}}
	\newcommand{\alphaturb}{\ensuremath{\alpha_\mathrm{turb}}}
	\newcommand{\tstar}{\ensuremath{T_*}}

	\newcommand{\chiism}{\ensuremath{\chi_\mathrm{ISM}}}
	\newcommand{\fpah}{\ensuremath{f_\mathrm{PAH}}}

	\newcommand{\mr}[1]{\ensuremath{\mathrm{#1}}}  

	\newcommand{\avgtgas}{\ensuremath{\langle \tgas \rangle}}
	\newcommand{\avggasdust}{\ensuremath{\langle \mathrm{g}/\mathrm{d} \rangle}}
	\newcommand{\tautwenty}{\ensuremath{\langle \tau _{20~\micron} \rangle}}
	
	\usepackage{placeins}
	\usepackage[arrows=pgf]{mhchem}
	
	\newcommand{\cem}[1]{\mbox{\ensuremath{\ce{#1}}}}

	\usepackage[utf8]{inputenc}
	\usepackage{pgfplots}
	\usepackage{grffile}
	\pgfplotsset{compat=newest}
	
	\pgfplotsset{yminorticks=true, major tick style={thick,black}, axis line style={thick,black}, minor tick style={semithick,black}, every axis plot/.append style={thick},every axis plot/.append style={scaled y ticks = false},y tick label style={/pgf/number format/fixed}} 
	
	\usetikzlibrary{plotmarks}
	\usetikzlibrary{arrows.meta}
	\usepgfplotslibrary{patchplots}
	\usepackage{amsmath}
	\newlength\figureheight
	\newlength\figurewidth
	\newlength\subheight
	\newlength\subwidth
	\newlength\subgap
	
	\usepackage[T1]{fontenc}
	\usepackage[utf8]{inputenc}
	\usepackage{pgfplots}
	\usepackage{grffile}
	\pgfplotsset{compat=newest}
	\usetikzlibrary{plotmarks}
	\usetikzlibrary{arrows.meta}
	\usepgfplotslibrary{patchplots}
	\usepackage{amsmath}
	\pgfplotsset{yminorticks=true, major tick style={thick,black}, axis line style={thick,black}, minor tick style={semithick,black}, every axis plot/.append style={thick}}

\begin{document}	
		
	\crefname{section}{Sect.}{Sects.}
	\crefname{table}{Tab.}{Tabs.}
		
		\title{The effects of dust evolution on disks in the mid-IR}
		
		\author{A.J. Greenwood$^1$
			\and
			I. Kamp$^1$ \and L.B.F.M. Waters$^{2,3}$ \and P. Woitke$^4$ \and W.-F. Thi$^5$
		}
		
		\institute{$^1$ Kapteyn Astronomical Institute, University of Groningen, Postbus 800, 9700 AV Groningen, The Netherlands\\
			$^2$ SRON Netherlands Institute for Space Research, P.O. Box 800, 9700 AV Groningen, The Netherlands \\
			$^3$ Anton Pannekoek Institute for Astronomy, University of Amsterdam, PO Box 94249, 1090 GE Amsterdam, The Netherlands \\ 
			$^4$ SUPA, School of Physics \& Astronomy, University of St. Andrews, North Haugh, St. Andrews KY16 9SS, UK \\
			$^5$ Max Planck Institute for Extraterrestrial Physics, Gie\textbeta{}enbachstra\textbeta{}e 1, 85741 Garching, Germany \\
			             \email{greenwood@astro.rug.nl}}

		\abstract{
In this paper, we couple together the dust evolution code \twopoppy{} with the thermochemical disk modelling code \prodimo{}. We create a series of thermochemical disk models that simulate the evolution of dust over time from $0.018~\myr$ to $10~\myr$, including the radial drift, growth, and settling of dust grains.  We examine the effects of this dust evolution on the mid-infrared gas emission, focussing on the mid-infrared spectral lines of \cem{C2H2}, \cem{CO2}, \cem{HCN}, \cem{NH3}, \cem{OH}, and \cem{H2O} that are readily observable with \spitzer{} and the upcoming \eelt{} and \jwst{}.

The addition of dust evolution acts to increase line fluxes by reducing the population of small dust grains. We find that the spectral lines of all species except \cem{C2H2} respond strongly to dust evolution, with line fluxes increasing by more than an order of magnitude across the model series as the density of small dust grains decreases over time.  The \cem{C2H2} line fluxes are extremely low due to a lack of abundance in the infrared line-emitting regions, despite \cem{C2H2} being commonly detected with \spitzer{}, suggesting that warm chemistry in the inner disk may need further investigation.
Finally, we find that the \cem{CO2} flux densities increase more rapidly than the other species as the dust disk evolves.  This suggests that the flux ratios of \cem{CO2} to other species may be lower in disks with less-evolved dust populations. 
}
	
	\maketitle

		\section{Introduction}
		
		Throughout the  few-million-year lifetime of a protoplanetary disk, its evolution is dominated by the gas and dust which accrete onto the star, coalesce into larger bodies, or are dissipated by photoevaporation and stellar winds. The dust in these disks is a significant factor in their evolution. Although dust particles (assuming typical ISM abundances) make up only a small portion of the total disk mass, small grains composed of dust and ice dominate the radiative transfer of the disk and provide ``seeds'' for planets to form.
		
		It is crucial that we can determine the distribution of this dust. In the eras of \alma{}, \jwst{}, and \eelt{} we have both the sensitivity and the angular resolution to spatially resolve nearby protoplanetary disks, to spectrally resolve the emission lines from molecules, and to detect the broad emission features from silicate grains. The distributions of gas and dust in the disk are not necessarily co-spatial, and

nor is the ratio constant with height, because dust grains are expected to grow and fall towards the midplane \citep{Blum:2008fi,Dominik:2008ic} and the scale heights of the dust and gas disks may differ \citep{deGregorioMonsalvo:2013jp,Avenhaus:2018tt}. T~Tauri stars and brown dwarfs have flatter dust disks and relatively thicker gas disks than Herbig stars \citep{Mulders:2012ew}. A crucial step that we must take is to understand the physics behind these effects, and how we can incorporate them into a thermochemical disk model.
		
		In the infrared, we can observe the spectral lines of many species that trace the warm upper layers of the inner disk. For example, typical mid-infrared water line fluxes in T~Tauri disks have been observed to range between about $2\times 10^{-13}~\erg~\mathrm{cm}^{-2}~\persec$ and $2\times 10^{-15}~\erg~\mathrm{cm}^{-2}~\persec$ \citep{Pontoppidan:2010gw}. The authors also note that there appears to be a sub-class of  disks which have strong $14.98~\micron$ \cem{CO2} emission, but with no detectable contributions in the $14$ to $16~\micron$ range from water, \cem{HCN}, and \cem{C2H2}.
		Previous results by \cite{Meijerink:2009jc} and \cite{2017AA...601A..36B} suggest that a high gas-to-dust ratio of $1000{:}1$ in the line-emitting regions might be necessary to reproduce the mid-infrared line fluxes that we have observed with \spitzer{}. This ratio has often been adopted by the modelling community \citep{Bruderer:2015iw,2017AA...601A..36B}.
		
		To justify this ratio, we need to explain what has happened to the dust: if the disk formed from a molecular cloud with a gas-to-dust ratio of $100{:}1$, this dust cannot disappear without reason. We know that dust settles in disks \citep{Dubrulle:1995jn}, and that especially in older disks, the midplanes are typically more dust-rich than the upper layers \citep{Dullemond:2005hf}. We need to account for this dust settling in order to quantify the gas-to-dust ratio of the line-emitting regions. In this paper, we couple a dust evolution code with a thermochemical modelling code and for the first time analyze the effects of dust evolution on the mid-infrared spectral lines.
		
		Accurately determining the gas and dust masses of protoplanetary disks is difficult. Primarily, because much of the line and continuum emission is optically thick: there is no known way to derive gas and dust masses from line and continuum observations that is free of significant degeneracies. Although observing disks with gaps can help to measure scale heights and thus estimate midplane densities, there can be a lot of optically-thick gas and dust in the midplane that cannot be observed directly -- for example, \citet{Pinte:2016kn} find that the dust density in the gaps of HL Tau is at least a factor of 10 lower than in the rings surrounding the gaps.  However, it is difficult to place an upper limit on the dust density in the rings.
		
		Likewise, it is difficult to accurately observe the dust grain size distribution as a function of radius and height.  	For example, although there is progress to spatially resolve the dust surface density of disks \citep{Pinilla:2014dl}, the measured spectral index is dependant upon assumptions about the dust temperature. Scattered light imaging using SPHERE has been used to observe the surface of dust disks \citep{Ginski:2016hy,Stolker:2016kg,Avenhaus:2018tt,MuroArena:2018ju}. Particularly, the survey of eight T~Tauri disks introduced by \cite{Avenhaus:2018tt} is a valuable set of observations. Although a detailed analysis of each source is left for forthcoming papers, \cite{Avenhaus:2018tt} provide insights such as showing that all of the observed T~Tauri disks appear to have $\tau=1$ surface flaring indices between $\sim 1.1$ and $\sim 1.6$, and shows  striking geometric structures such as gaps, rings, and even the lower disk surface of three disks. If an inclined disk has rings or gaps in the dust structure, the height of the dust scattering surface can be estimated by geometrically fitting these structures  \citep{Ginski:2016hy,Avenhaus:2018tt}. 
		
		In comparison to micron-sized dust grains, decimetre-sized bodies have very little effect on the radiative transfer of the disk in the mid-infrared regions. However, their effect on the disk's evolution is much more significant. Because they are less affected by processes such as viscosity and turbulence, decimetre-sized bodies are more likely to be found close to the disk's midplane, and are a crucial step towards the eventual formation of planets.

To analyze the effect of dust evolution and migration on the mid-infrared spectral lines of a T~Tauri protoplanetary disk, we model the evolution of dust over time, and create a thermochemical model of the disk at discrete time steps in the evolution. In this way, we incorporate a self-consistent description of dust evolution into our thermochemical disk models, where the gas-to-dust ratio and dust size distribution vary not only at each timestep, but also  radially and vertically through the disk.  By analyzing the mid-infrared spectral lines of each individual model, we can build an understanding of how dust evolution affects these lines.

		\section{Modelling strategy}

		A total of four separate codes were used in a set order to produce the models in this paper. The underlying disk model is based upon the ``DIANA standard'' T~Tauri disk described in \citet{Woitke:2016gp}, except for the dust structure.  \Cref{tab:modelparams} describes the parameters of our T~Tauri disk model. This section serves as a technical description of what each code does and how they are interfaced.
		
		First, we use the two-population dust evolution code \twopoppy{}\footnote{\twopoppy{} is available on Github at \url{http://birnstiel.github.io/two-pop-py/}.} \citep{Birnstiel:2012ft,Birnstiel:2015cl,Birnstiel:2017vr}. The \twopoppy{} code produces a one-dimensional description of the dust at each age increment. It is a simplified version of the dust evolution code described in \citet{Birnstiel:2010eq}, calibrated with the use of these more complex models and based upon the use of two dust grain populations: a small monomer population, and a large grain population. The small grain population is coupled to the gas, not affected by drift velocities, and is constant in time and space. The  population of large grains can grow in size, and is affected by the limiting mechanisms of radial drift and grain fragmentation. The reconstruction of the grain size distribution from the simple two-population model is described by \citet{Birnstiel:2015cl}. 
		
		From \twopoppy{}, we obtain the gas and dust surface densities as functions of radius, and the dust mass fraction as a function of both radius and grain size  (which records the fraction of the total mass at each radius that is accounted for by each grain size). For each age increment, we calculate a separate disk model. For this, we use the code \mcmax{} \citep{Min:2008aj} to create a two-dimensional model of the disk structure and calculate the dust continuum radiative transfer for the dust temperature. We ensure that the grain sizes are conserved between \twopoppy{} and \mcmax{}.
		
		There are two standard methods in \mcmax{} to calculating the scale heights of the gas and dust: to have parameterized scale heights, or to solve for hydrostatic equilibrium. We take a hybrid approach, first forcing the scale heights of the gas to remain parameterized, then solving the dust for hydrostatic equilibrium and allowing the dust to settle in a self-consistent manner (described by \citealt{Mulders:2012ew}). We do not argue for the physical correctness of this procedure (which is unknown), rather for the simplicity it brings. The first argument is that we want to remain as close to the DIANA standard of models as possible, where only the dust distribution deviates from the standard. The second argument is that it simplifies the results and analysis: although the total gas mass in the disk decreases with age, the gas scale heights remain constant. This allows us better to analyse the effects that different dust distributions have on the mid-infrared molecular lines, under the assumption that everything else is equal.

The third code in the pipeline is \prodimo{} \citep{Woitke:2009jf,Kamp:2010ek,Aresu:2011cm}, which takes the \mcmax{} output and calculates a full thermochemical disk model. We take the disk structure directly from \mcmax{}, then run the disk chemistry, taking the reaction rates for 235 species from the UMIST2012 chemical network \citep{McElroy:2013ki} to calculate the disk's thermal and chemical equilibrium state. Because the overall opacity and density structure of the disk is fixed, there is no need for global disk structure iterations.
	
		The final element is \flits{}, which calculates a high-resolution spectrum from a \prodimo{} model  \citep{Woitke:2018ux} across the infrared wavelength range. \prodimo{} provides the gas and dust temperatures, the stellar spectrum and local radiation field, opacities, molecular number densities, and level populations. \flits{} reads this file and calculates the disk's spectrum  for the desired molecules and wavelength range.

		\begin{table*}
			
			\footnotesize
			\caption{\footnotesize A summary of important model parameters used in each step of the procedure. Some parameters, for example the disk mass, are important for \twopoppy{}, \mcmax{}, and \prodimo{}. These parameters are not listed for the subsequent codes because their effects are embedded in the data passed between the codes. The only parameter in this table that is passed through to \flits{} is the disk inclination. Of the \twopoppy{} parameters in this table, the disk mass and gas-to-dust ratio are only valid for the initial state of the disk. The other parameters in this table are constant in time. \ngrains{} is the number of grain sizes used for \twopoppy{}'s reconstruction of the grain size distribution.
			The dust is a distribution of hollow spheres, where the maximum fractional volume filled by the central void is 0.8 \citep{Min:2005uy,Min:2016hr}.
			The dust grain mixture in \mcmax{} is $60\%$ amorphous \cem{Mg_{0.7} Fe_{0.3} Si O_3} silicates \citep{Dorschner:1995wq}, $15\%$ amorphous carbon \citep{Zubko:1996fn}, and $25\%$ vacuum.  The viscosity helps define the surface density $\Sigma_\mathrm{g}(r)$, where $\Sigma_\mathrm{g}(r) \propto \left( {r}/{r_\mathrm{c}} \right) ^{-\gamma_{\mathrm{visc}}}  \exp \left[ - \left( {r}/{r_\mathrm{c}}\right) ^{2-\gamma_{\mathrm{visc}}} \right] $ with a characteristic radius $r_\mathrm{c}=60~\au$ \citep{Birnstiel:2012ft}.
}
			\centering 
			\begin{center}
			\begin{tabular}{l l l l l}
				Symbol & Quantity & \twopoppy	& \mcmax & \prodimo \\ \midrule
				\rstar & Stellar radius & $2.086~\rsun$ & $2.086~\rsun$ \\
				\mstar & Stellar mass & $0.70~\msun$  & $0.70~\msun$ & $0.70~\msun$ \\
				$\mdisk / \mstar$ & Initial disk mass & $0.1$  \\
				\rtaper & Taper radius & $100~\au$ & $100~\au$ \\
				$\gamma_{\mathrm{visc}}$ & Viscosity exponent & $1$ \\
				\ngrains & Number of dust grains & $150$ & $150$ \\
				\amin & Minimum dust grain size &   $ 10^{-5}~\cm$ &  $ 10^{-5}~\cm$ \\
				\amax & Maximum dust grain size &  & \multicolumn{1}{l}{$1.9\times 10^2~\cm$} \\
				$f_\mathrm{vac}$ & \multicolumn{1}{l}{Dust grain porosity (vacuum fraction)} & $0.25$ \\
				$f_\mathrm{max}$ & \multicolumn{1}{l}{Maximum hollow volume ratio} & & $0.8$ \\ \vspace{0em}
				\tmax & Final timestep  & $10^7$ years \\
				\alphaturb & Turbulence & $10^{-3}$ & $10^{-3}$ \\
				\rin & Inner radius & $0.06835~\au$ & $0.07~\au$  \\
				\rout & Outer radius & $2000~\au$ & $600~\au$ \\
				$g/d$ & Initial gas-to-dust ratio & $100$ \\
				\tstar & Stellar surface temperature & $4000~\kelvin$ & $4000~\kelvin$ & $4000~\kelvin$ \\
				\nrad & Number of radial grid points & $240$ & $500$ & $240$ \\
				\ntheta & Number of azimuthal grid points & & $150$ \\
				$N_\mathrm{Z}$ & \multicolumn{1}{l}{Number of vertical grid points} & & $160$ \\
				$N_\mathrm{inner}$ & \multicolumn{1}{l}{Number of radial points near inner rim} & & $100$ \\
				\texttt{gasevol} & Gas evolution & \texttt{true} \\
				$D$ & Distance of the disk & & $140~\au$ & $140~\au$ \\
				$i$ & Inclination of the disk  & & $45^\circ$ & $45^\circ$ \\ 
				\chiism & \multicolumn{1}{l}{Interstellar radiation field \citep{Draine:1978ec}} & $1$ & $1$ \\
				$H_0$ & Gas scale height at $1~\au$ & & $0.05012~\au$ \\
				$\beta$ & Flaring exponent & & $1.15$ \\
				\texttt{scset} & Self-consistent dust settling & & \texttt{true} \\
				\lstar & Stellar luminosity & & & $1.0~\lsun$ \\
				\fpah & \multicolumn{1}{l}{PAH abundance (relative to ISM)} & & $10^{-4}$ \\
				$\fuv / \lstar$ & UV excess & & & $0.01$ \\
				\puv & UV powerlaw exponent & & & $1.3$ \\
				$\zeta_\mathrm{CR}$ & \multicolumn{1}{l}{Cosmic ray \cem{H2} ionization rate} & & \multicolumn{1}{l}{$1.7\times 10^{-17}~\persec$} \\
				\lxr & X-ray luminosity & & & \multicolumn{1}{l}{$10^{30}~\erg~\persec$} \\ \midrule
			\end{tabular}
		\end{center}
			\label{tab:modelparams}
		\end{table*}

		\section{Dust migration and disk surface densities}
		
		The effects of dust migration over time are visible in both the surface densities and dust grain size distributions, as computed by \twopoppy{}. These are one-dimensional results, while we later use \mcmax{} to compute the two-dimensional disk structure. The precise timescales are not strictly relevant for this discussion because the ages are very uncertain, both in the models and any observed disks. It is better to compare disks with different surface densities than to compare disks of different ages, because these parameters are more readily observable. For example, disks in Ophiuchus are thought to be less than $1~\myr$ old \citep{Furlan:2009ci}, but basic properties such as the disk mass may vary by about two orders of magnitude \citep{Testi:2016tw}. We need many different models in order to accurately represent such a diverse disk population. Nevertheless, for brevity we refer to each model by its \twopoppy{} age. 
		
		In each \prodimo{} model we use steady-state chemistry: the  \prodimo{} chemistry may run for longer than the ``age'' of the \twopoppy{} model. However, the timescales for disk surface chemistry (on the order of years) are much shorter than the timescale of dust evolution ($\gtrsim 10^4~\mathrm{years}$): it is only when considering longer wavelengths such as the sub-mm that we might need to calculate time-dependent chemistry.
		
\begin{figure}                          
		\centering                                                                                                                                                                                                                                
		\includegraphics{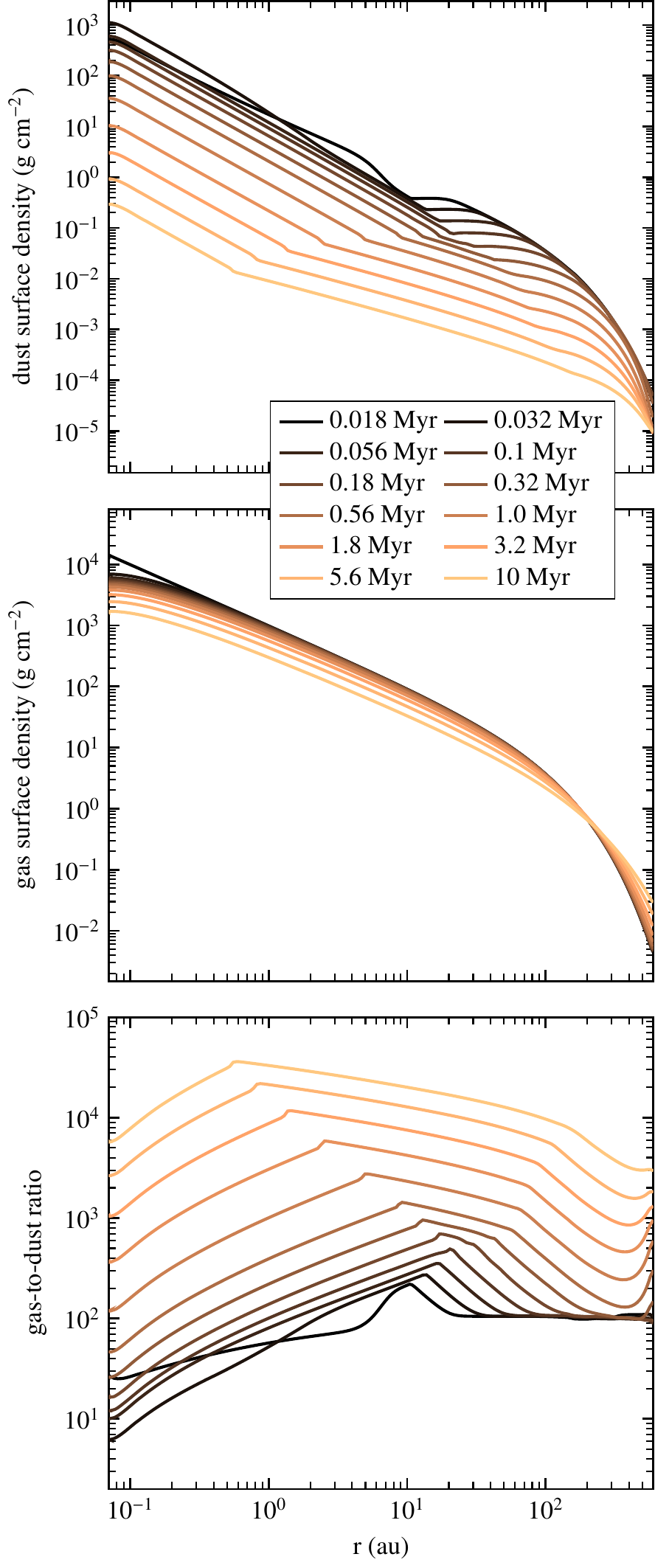}
		\caption{The dust surface column density, the gas surface column density, and the gas-to-dust ratio as a function of radius, for each age increment.}
		\label{fig:radial_properties_combined}                                                                                                                                                                                                                                         
	\end{figure}

		\Cref{fig:radial_properties_combined} show the gas and dust surface densities from $t_0 = 0.018~\myr$ to $\tmax = 10~\myr$.  Although the gas mass does slowly decrease over time due to viscous evolution, the loss of dust due to radial drift is much greater\footnote{To simplify the analysis of comparing different dust structures, we do not include the accretion luminosity in our \mcmax{} and \prodimo{} models.}, and this causes the total gas-to-dust ratio of the disk to increase over time, from $100$ at $t=0$ to over $6000$ at $10~\myr$. For the column of gas at a radius of $1~\au$, the youngest model has a gas-to-dust ratio of about $40$, whereas the oldest model reaches about $4\times 10^4$.

		The youngest \twopoppy{} model is close to the initial state of the system, and the low gas-to-dust ratio in the initial state of the inner disk is the result of a high level of inwards dust mass flux. As the disk ages, radial drift continues to deplete the total dust mass in the disk. One other notable trend in \cref{fig:radial_properties_combined} is the ``bump'' in gas-to-dust ratio, which becomes significant at $0.32~\myr$ and a radius of $40~\au$, and moves inwards with time. This bump is due to a localized depletion in the dust, reported by \citet{Birnstiel:2012ft} as a pile-up effect caused by increases in the disk temperature due to viscous heating, similar to the grain pile-up in \citet{Youdin:2004ff}.

\begin{figure}
			\centering
			\includegraphics{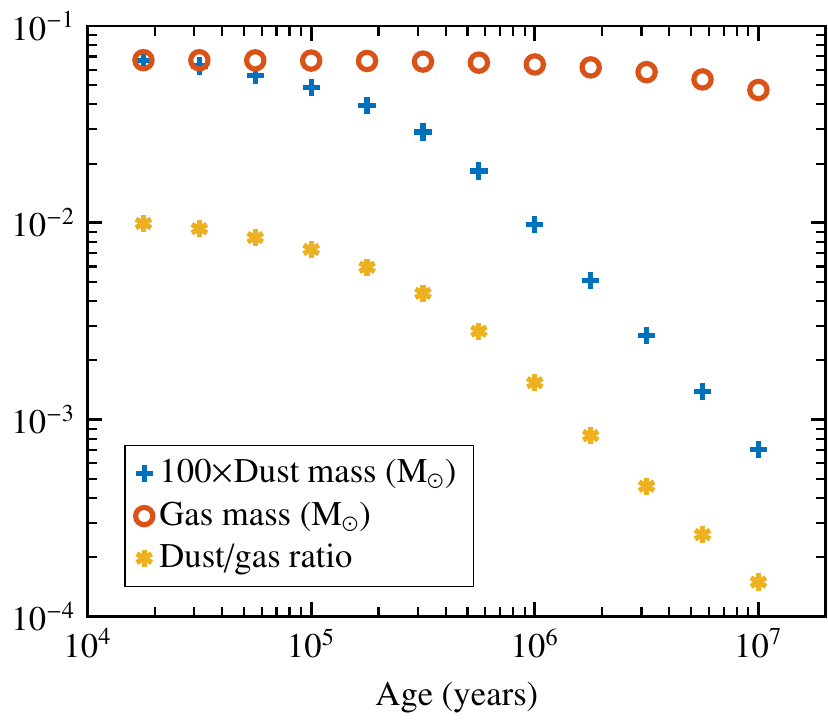}
			\caption{The dust and gas masses and dust-to-gas ratio of the entire disk over time.}
			\label{fig:dustev_gasdustmass_ratio}
		\end{figure}

\begin{figure*}
					\centering
					\includegraphics{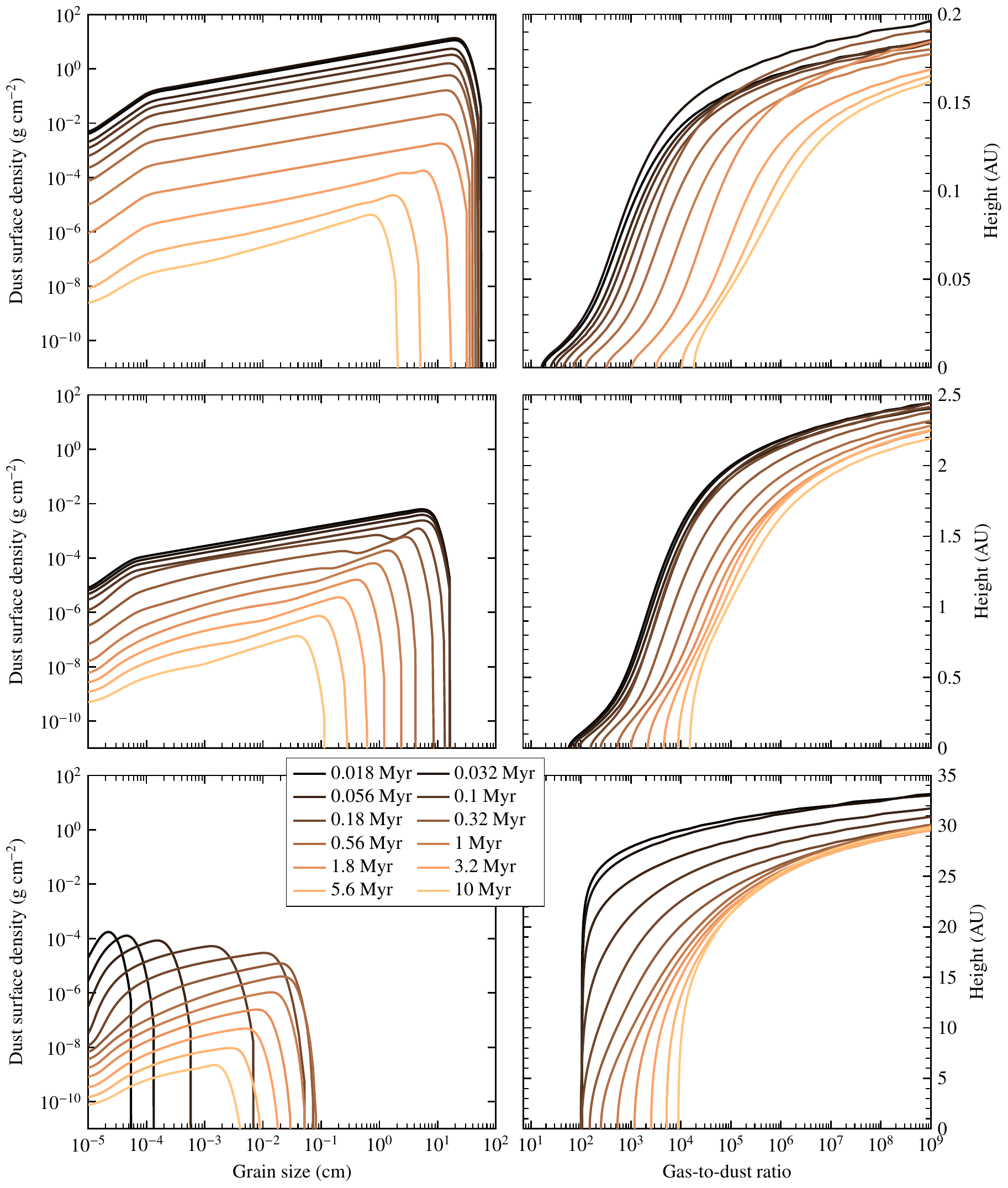}
					\caption{\textbf{Left:} Vertical cuts of the grain size distribution at $r=1~\au$ (top), $r=10~\au$ (middle), and $r=100~\au$ (bottom). \textbf{Right:} vertical cuts of the gas-to-dust ratio  at $r=1~\au$ (top), $r=10~\au$ (middle), and $r=100~\au$ (bottom).}
					\label{fig:vercut_combined}
				\end{figure*}

		\Cref{fig:dustev_gasdustmass_ratio} describes how the gas mass, dust mass, and gas-to-dust ratio vary over time, integrated over the entire disk.  \Cref{fig:vercut_combined} shows vertical cuts of the grain size distribution at radii of $0.1$, $1$, $10$, and $100~\au$.  At radii of $10~\au$ or less, the trends are similar: the grain size distribution shows both an overall decrease in surface densities, and a disproportionate depletion of larger grains (which are transported inwards due to radial drift). As a consequence, in the inner disk at $0.1~\au$ there are many large grains up to $10~\cm$ in size at an age of $10~\myr$. However, at $1~\au$ this cut-off is at a grain size of $1~\cm$, and about $0.05~\cm$ at $10~\au$.  This happens because a population of large grains can only be maintained so long as there is an inwards flux of grains coming from larger radii. The situation is slightly different at $100~\au$: although there is no significant population of cm-sized, grains, the population of mm-sized grains grows until an age of $0.56~\myr$.

		\section{Size of the dust disk}
		
		One consequence of our dust settling method is that the dust disk is less vertically-extended than the gas disk. This is not an artefact of parameterizing the gas structure, but a result of the self-consistent description of dust settling.  Disks with processed dust and dust disks that are thinner than the gas disk have been reported in previous literature  \citep{deGregorioMonsalvo:2013jp,Ginski:2016hy,MuroArena:2018ju,Avenhaus:2018tt}, but the number of observations is still relatively few.
		
		The dust disk calculated using our self-consistent settling is thinner than a disk model that uses Dubrulle settling \citep{Dubrulle:1995jn}.  The main differences between the self-consistent settling method  \citep{Mulders:2012ew} and Dubrulle setting is that Dubrulle settling uses only the midplane density and temperature. Away from the midplane, their analytical solution becomes less accurate. Self-consistent settling agrees well with Dubrulle settling in the midplane, but sub-micron particles can also decouple from the gas in the upper disk   \citep{Mulders:2012ew}, contrary to Dubrulle settling. The distribution of $<0.1~\micron$ particles is nearly uniform throughout a Dubrulle-settled disk, and larger grain sizes are affected less by settling than the self-consistent model  \citep{Mulders:2012ew}.

		All of our models have a dust disk that is thinner than the gas disk. \Cref{fig:vercut_combined} shows vertical cuts of the gas-to-dust ratio through the disks at each age: in every case, in the upper layers of the disk where some tenuous gas still exists, the gas-to-dust ratio can easily be $10^5$ or more. In the midplane of the inner disk, particularly at $0.1~\au$, the gas-to-dust ratio of models younger than $0.56~\myr$ can be less than $10$. This is in contrast to at $100~\au$, where most of the disk at $0.018~\myr$ has a gas-to-dust ratio of $100$. At young ages, there are few large grains in the outer disk: the dust only decouples significantly from the gas in older models, once mm-sized dust grains have formed. However, at small radii, there is already a significant population of large grains at $t=0.18~\myr$. This gives rise to the low gas-to-dust ratio in the midplane at small radii.
		
When comparing the  gas and dust disk thickness with observations, our youngest models are comparable to disks such as IM Lup that appear to be less settled \citep{Cleeves:2016fi}, while the older models are more comparable to disks such as HD~163296, where the dust disk may be significantly settled in comparison to the gas disk \citep{GregorioMonsalvo:2013}.
		
		The  settled dust structure causes every disk to have some regions where the gas-to-dust ratio is $1000$ or more, which as we find in \cref{sec:mid-irspec} is the minimum gas-to-dust ratio from which all species except \cem{C2H2} tend to emit.
		
\begin{figure}
			\centering
			\includegraphics{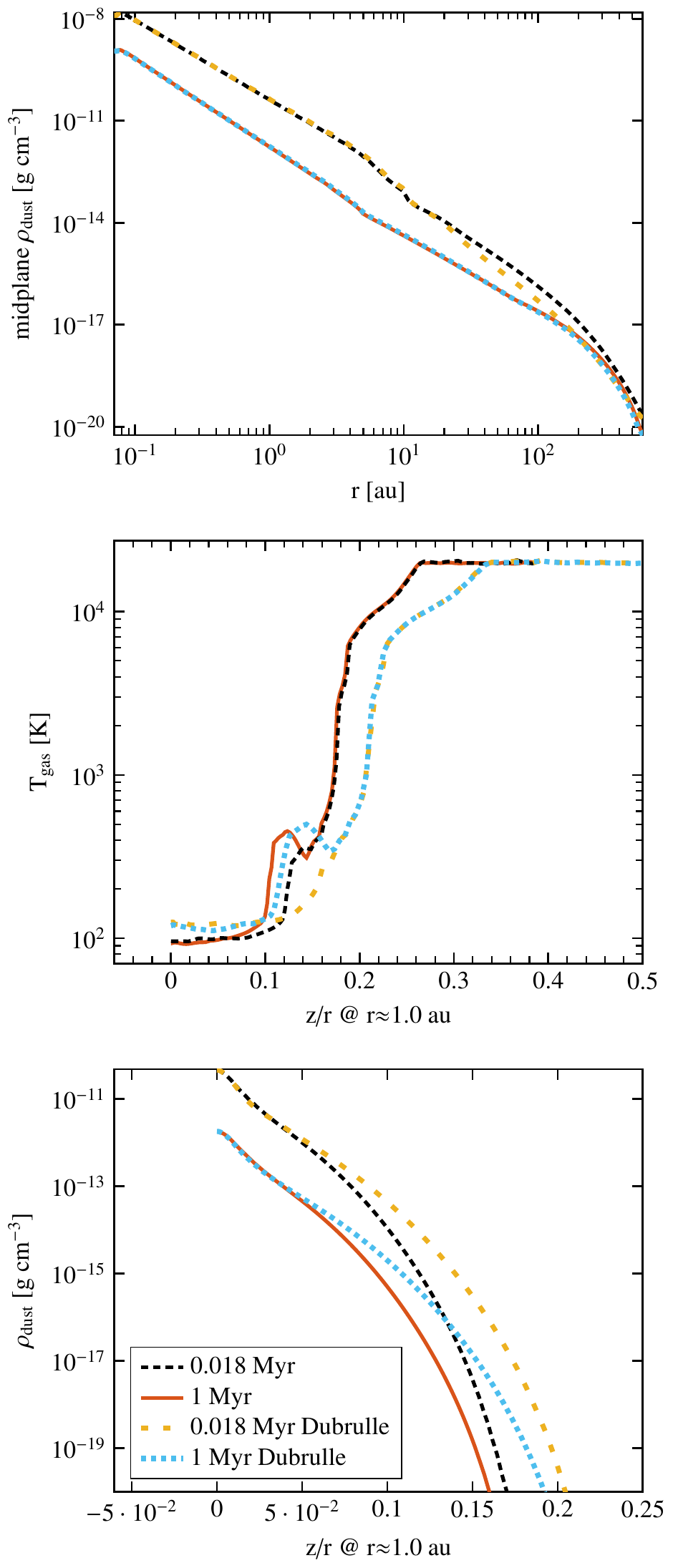}
			\caption{From top to bottom:  the midplane dust density, gas temperature of a vertical column of gas at $1~\au$, and a vertical cut of the dust density at $1~\au$.  For comparison with the standard $0.018~\myr$ and $1~\myr$ models, the same models (but with Dubrulle settling instead) have been plotted, in order to illustrate the effects of both dust settling and evolution on the mid-infrared regions.}
			\label{fig:evol_surfdens_comparison}
		\end{figure}

\begin{figure}
			\centering
			\includegraphics{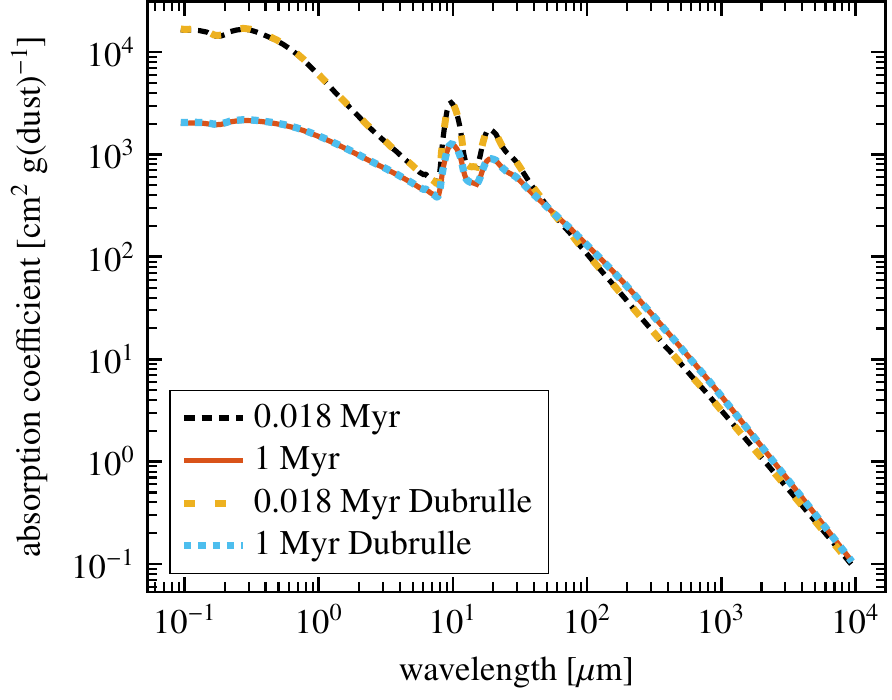}
			\caption{The absorption opacity, plotted for the standard $0.018~\myr$ and $1~\myr$ models, and the same models but with Dubrulle settling instead.}
			\label{fig:absorptioncoefficient}
		\end{figure}

\begin{figure}
			\centering
			\includegraphics{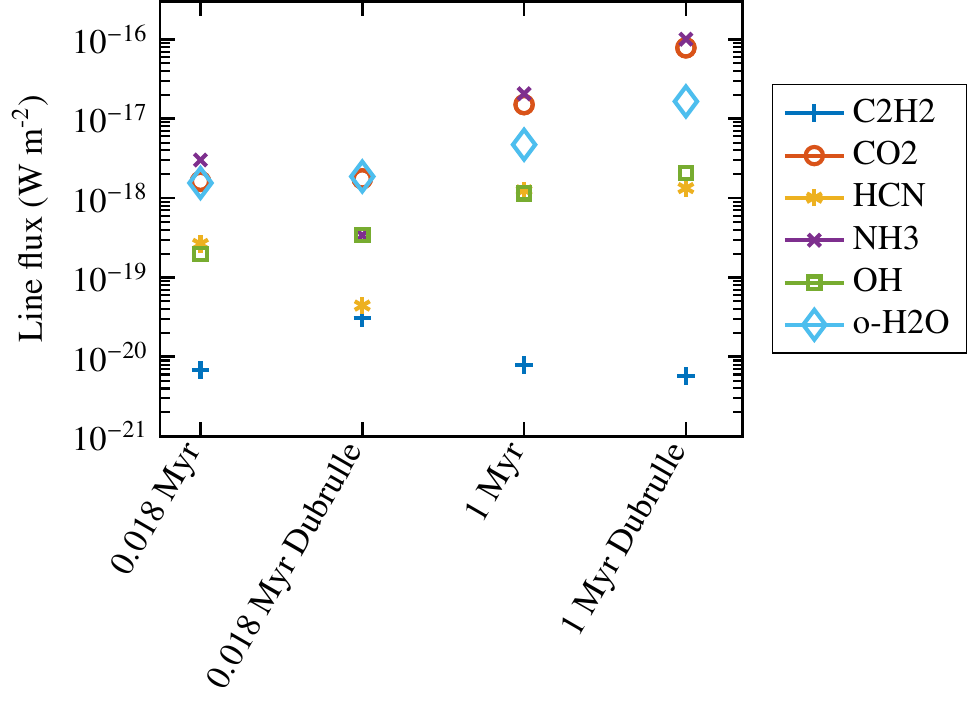}
			\caption{Mid-infrared line fluxes for both the standard $0.018~\myr$ and $1~\myr$ models, and the same models with Dubrulle settling.}
			\label{fig:lineflux_perspecies_dubrulle}
		\end{figure}

			Both dust evolution and the new self-consistent method of settling contribute to changes in the mid-infrared line fluxes, because both of these processes affect small micron-sized dust grains.  Dust evolution affects the surface density of these grains in the disk, while settling affects the vertical distribution of those grains.  \Cref{fig:evol_surfdens_comparison,fig:absorptioncoefficient} shows the effects of dust evolution and the settling method on dust densities, absorption coefficient, and gas temperatures, while \cref{fig:lineflux_perspecies_dubrulle} shows those same effects on the mid-infrared line fluxes.
		
		Counter-intuitively, the mid-infrared line fluxes can in fact increase due to Dubrulle settling. Even though Dubrulle settling leads to lower gas temperatures in the mid-infrared line-emitting region for each species, in \cref{fig:lineflux_perspecies_dubrulle} we do see that models with Dubrulle settling tend to have somewhat stronger lines than those same models with self-consistent settling.  \textbf{Although the total vertical optical depth remains the same in models with both Dubrulle and self-consistent settling, the dust in the latter models is more concentrated  towards the midplane. Thus, reduced optical depths in the upper layers of models with Dubrulle settling can allow the mid-infrared line-emitting regions to grow in size. However, it is important to note that although the method of settling can affect the line fluxes, the effects of dust evolution remain more significant.}

		\section{Mid-infrared spectra}
		\label{sec:mid-irspec}
		
		The migration and evolution of dust has a profound effect on the mid-infrared spectra and line-emitting regions of our disk model. To discuss the line-emitting regions of each species, it is necessary  to define the exact molecular lines we are using, and how the line-emitting regions themselves are derived.
		
		We define the line-emitting region as the area from which $70\%$ of the flux originates, in both the radial and vertical directions. The lower limit, $\mr{x15}$, is defined so that $85\%$ of the total line flux of that spectral line is emitted at radii greater than $\mr{x15}$. The upper limit, $\mr{x85}$, is the opposite -- only $15\%$ of the total line flux of that spectral line is emitted at a radius greater than $\mr{x85}$. 
At each radius, $\mr{z15}$ and $\mr{z85}$ define the heights in the disk above which $15\%$ and $85\%$ of the flux is emitted respectively. The properites of the line-emitting region for a given species (such as \tgas)  are averaged and weighted with the volume density of that species across the line-emitting region.
		
\Cref{tab:transitions_dustpaper} details the molecular lines chosen for analysis, one for each species. For consistency, where possible, we analyze spectroscopic lines that have previously been analyzed in other literature (see references in \cref{tab:transitions_dustpaper}). Whenever an individual molecular line is referenced in this paper, it refers to the line in this table.
		
		\begin{table*}
			\begin{center}
				\caption{The emission line of each species chosen for analysis, including upper level energies  $E_\mathrm{up}$ and the Einstein $A$ coefficient.
					The description of the ro-vibrational lines of \cem{CO2}, \cem{C2H2}, \cem{HCN}, and \cem{NH3} is an abbreviated form of that described in \citet{Jacquemart:2003bc,Rothman:2005cj}, where $v_j$  are the normal mode vibrational quantum numbers, $l_j$ are the vibrational angular momentum quantum numbers, and $l$ is the absolute value of the sum of $l_j$. The final entry, for example $R11e$, denotes that it is an $R$-branch transition, the lower-state rotational energy level is $11$, and $e$ or $f$ denotes the symmetry for $l$-type doubling.}
				\medskip
				\centering
				\begin{tabular}{l l l l l l}
					Species & $\lambda~(\micron)$ & Transition & $E_\mathrm{up}~(\kelvin)$ & $A~(\mathrm{s}^{-1})$ & Reference \\  \midrule 
					\cem{CO2} & $14.98299$ & $v_1v_2l_2v_3r = 01101 \rightarrow 00001$, $Q6e$ &  $983.85$ & $1.527$ & \cite{2017AA...601A..36B} \\
					\cem{C2H2} & $13.20393$ & $v_1v_2v_3v_4v_5l\pm = 000011 \rightarrow 000000$,  $R11e$ & $1313.1$ & $3.509$ & \citet{Woitke:2018ux} \\
					\cem{HCN} & $14.03930$ & $v_1v_2l_2v_3 = 0110 \rightarrow 0000$, $Q6e$ & $1114.1$ & $2.028$	& \cite{Bruderer:2015iw} \\
					\cem{o-H2O} & $17.75408$ & $J'=6 \rightarrow J''=5$ & $1278.5$ & $0.002869$ & \cite{Notsu:2017jc} \\
					\cem{NH3} & $10.33756$ & $v_1v_2v_3v_4 = 0100 \rightarrow 0000$, $J'=3 \rightarrow J''=3$ & $1515.3$ & $11.57$ &  \\
					\cem{OH} & $20.11506$ &$J'=13.5 \rightarrow J''=12.5$  & $5527.2$ & $50.47$ & \citet{Woitke:2018ux} \\ 
				\end{tabular}
				\label{tab:transitions_dustpaper}
			\end{center}
		\end{table*}

		The flux of a particular molecular line is driven primarily by three factors: the gas-to-dust ratio, the gas temperature in the line-emitting region, and the optical depth in the line-emitting region \tautwenty{}. \Cref{fig:perspecies_combined} shows how the average gas-to-dust ratio \avggasdust{}, average
		gas temperature \avgtgas{}, and the line fluxes change with age in the disk.  Although the gas-to-dust ratio of the entire disk begins at $100{:}1$, the self-consistent dust settling is not informed by the ``age'' of the model. It solves for the vertical dust structure without considering the settling timescales, which may be a few $\times 10^{5}$ years \citep{Dullemond:2004jj}. 
		
		We measure fluxes in two different ways.  Where we mention the flux of an individual spectral line, these fluxes are calculated using a vertical escape probability method \citep{Woitke:2009jf}. These line fluxes are accurate only for a face-on disk, because no detailed radiative transfer is taken into account.  In contrast, where discuss \flits{} spectra, these have been calculated for an inclined disk and account for both the radial and vertical optical depth. They are also convolved to an instrumental resolution, so that we measure the peak flux of a complex of lines and not the integrated flux of a single line.  The purpose of the first method is to measure how line emission responds to changes in the dust distribution, by picking these representative lines to measure the properties of the line-emitting region.  The purpose of the second method is to put these measurements into an observational context. By accounting for the many spectral lines that are in the bandheads of some species and convolving the resulting spectrum, we can relate the peak fluxes of the spectra to observations and sensitivities of \spitzer{} or \jwst{}.

In comparison to observed \spitzer{} spectra (e.g. \citealt{Pontoppidan:2010gw,Salyk:2011jz,Pascucci:2013te}), the \flits{} spectra of all of our species except \cem{C2H2} can produce spectral lines that are at least as bright as we have observed. The spectra are highly responsive to changes in the dust distribution, where the loss of small dust particles over time acts to reduce the opacity of the disk and dramatically increases line fluxes. The only exception to this trend is \cem{C2H2}, where our models do not produce lines that are as bright as \spitzer{} observations (despite \cem{C2H2} being readily detected), suggesting that the chemical networks of \cem{C2H2} may be inaccurate or incomplete.

		\subsection{Line fluxes: C$_2$H$_2$ in comparison to other species}
		
		Throughout the dust evolution simulation, the line flux of each species except \cem{C2H2} increases by at least an order of magnitude, whereas \cem{C2H2} remains more or less constant in line flux. 		
		For all species except \cem{C2H2}, the gas-to-dust ratio in the line-emitting region increases significantly as the dust evolves.
		
		On the other hand \cem{C2H2} simply does not exist in these models in the upper layers of the disk. It is easily destroyed when the dust evolves and the disk becomes more optically thin. As the disk ages the remaining \cem{C2H2} is concentrated further towards the midplane, and the line-emitting region is pushed both towards the midplane and out to larger radii (up to $\sim 0.6~\au$). In these  models, we see the same bifurcated structure of \cem{C2H2} discussed in {Greenwood et al. (submitted)} and seen also in \citet{Walsh:2015jr}. A significant proportion of the line-emitting population of \cem{C2H2} is below the $\av=10$ line, meaning that the dust continuum in this region is very optically thick in the mid-infrared. As the dust evolves and becomes more optically thin, the \cem{C2H2} responds and shifts closer towards the midplane. Hence, there is no chance for the line flux to increase. Greenwood et al. (submitted) show that a more dust-rich disk model has stronger \cem{C2H2} lines, where for all other species, more dust-rich models have weaker lines. Our models from this work do not predict \cem{C2H2} line fluxes that are observable with any current spacecraft. However, there are clear \spitzer{} detections of \cem{C2H2} in T~Tauri disks with similar detection rates to the other species \citep{Pontoppidan:2010gw}.  More work is needed to understand this particular species.

		\subsection{Properties of the line-emitting regions}

\begin{figure*}
					\centering
					\includegraphics{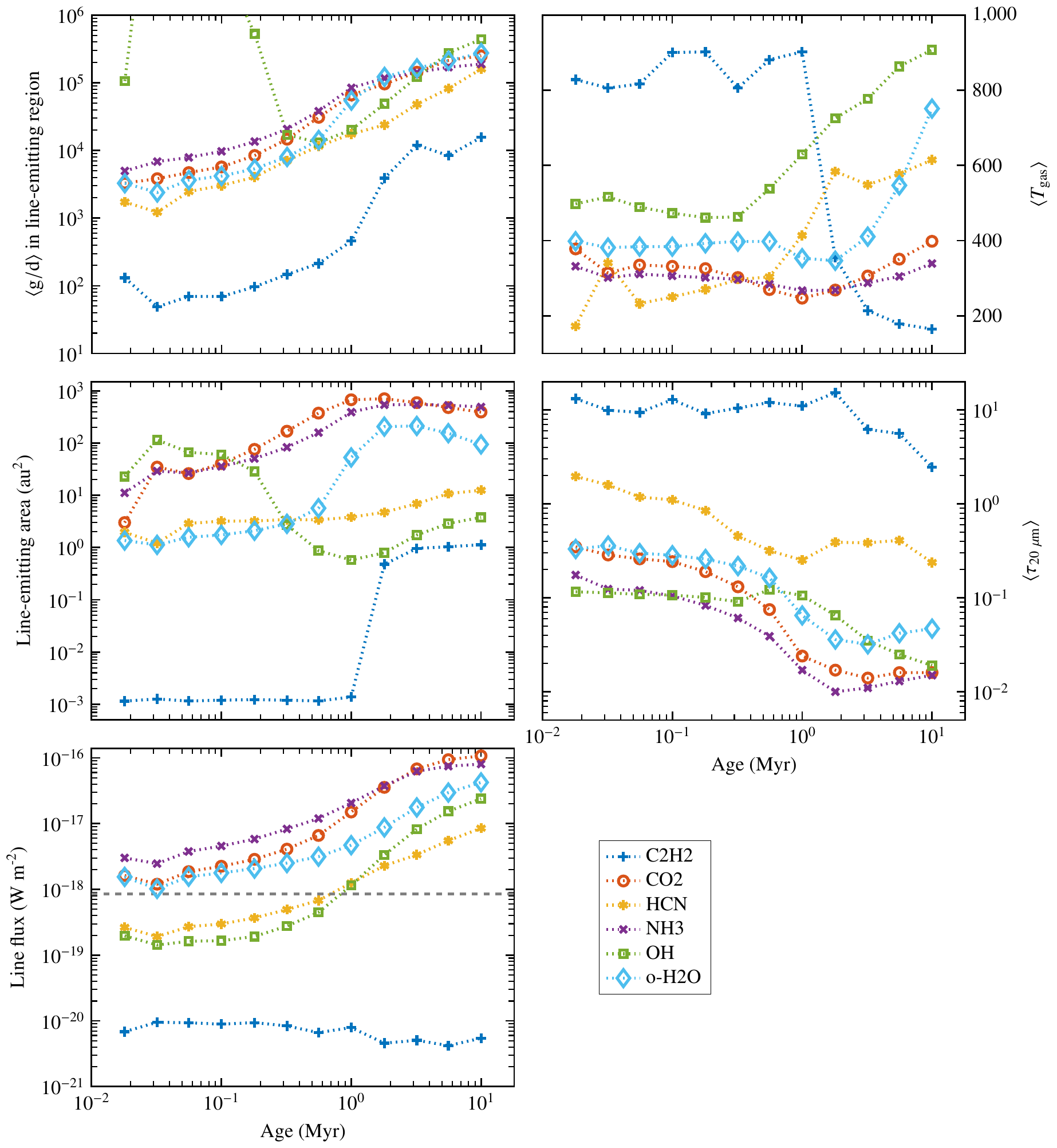}
					\caption{\textbf{Left, top:}  the gas-to-dust ratio in the line-emitting region of each species, over time. The significant bump seen in the gas-to-dust ratio of \cem{OH} is misleading: our measurement of the line-emitting area is biased by the fact that there is a small amount of \cem{OH} emission coming from warm upper layers at relatively large radii. These upper layers have a disproportionately high gas-to-dust ratio.
						   \textbf{Left, middle:} the line-emitting area, in square astronomical units, of each species over time. This area is calculated from a face-on perspective, assuming that the line emission comes from an annulus encompassed by the inner and outer radii of the line emission ($x15$ and $x85$). Much like the gas-to-dust ratio for less-evolved models, the line-emitting area calculation of \cem{OH} is biased by small amounts of emission at larger radii.
						   \textbf{Left, bottom:} the escape probability line flux of each species, over time.  The dashed horizontal line indicates the nominal $1~\sigma$ Spitzer sensitivity of $8.5\times 10^{-19}~\mathrm{W}~\mathrm{m}^{-2}$ achievable with a 512 second integration. 
						   \textbf{Right, top:}  the density-averaged gas temperature in the line-emitting region of each species, over time.
						   \textbf{Right, middle:} the density-averaged vertical optical depth of the dust at $20~\micron$ in the line-emitting region of each species, over time.
						   }
					\label{fig:perspecies_combined}
				\end{figure*}

		\Cref{fig:perspecies_combined} shows changes in the gas-to-dust ratio, gas temperature, vertical dust optical depth, escape probability line flux, and line-emitting area of each line over time. Although for each species the gas-to-dust ratio in the line-emitting region increases steadily over time, \cem{C2H2} has a consistently lower ratio than the other species because its line-emitting area is closer towards the mid-plane. 

		The trends in gas temperature are more complex. For \cem{C2H2}, at $1.8~\myr$ the gas temperature of the line-emitting region drops dramatically, from about $900~\kelvin$ to $350~\kelvin$. The temperature drops even more as the disk ages further. This trend is also reflected in the fact that \cem{C2H2} is in absorption for the oldest few disk models (see \cref{fig:flitsspec2}). The reason for this is changes in the line-emitting area: the \cem{C2H2} abundances in the oldest disk models drop significantly, and the line-emitting area moves outwards and towards the mid-plane.  The gas temperatures of the \cem{CO2} and \cem{NH3} line-emitting regions stay relatively stable over time. For \cem{HCN} and \cem{H2O}, we see an increase in gas temperature, particularly for the oldest models. This is because the line-emitting regions become larger and move towards slightly warmer, higher layers in the disk. 
		
\begin{figure*}
			\centering
			\includegraphics{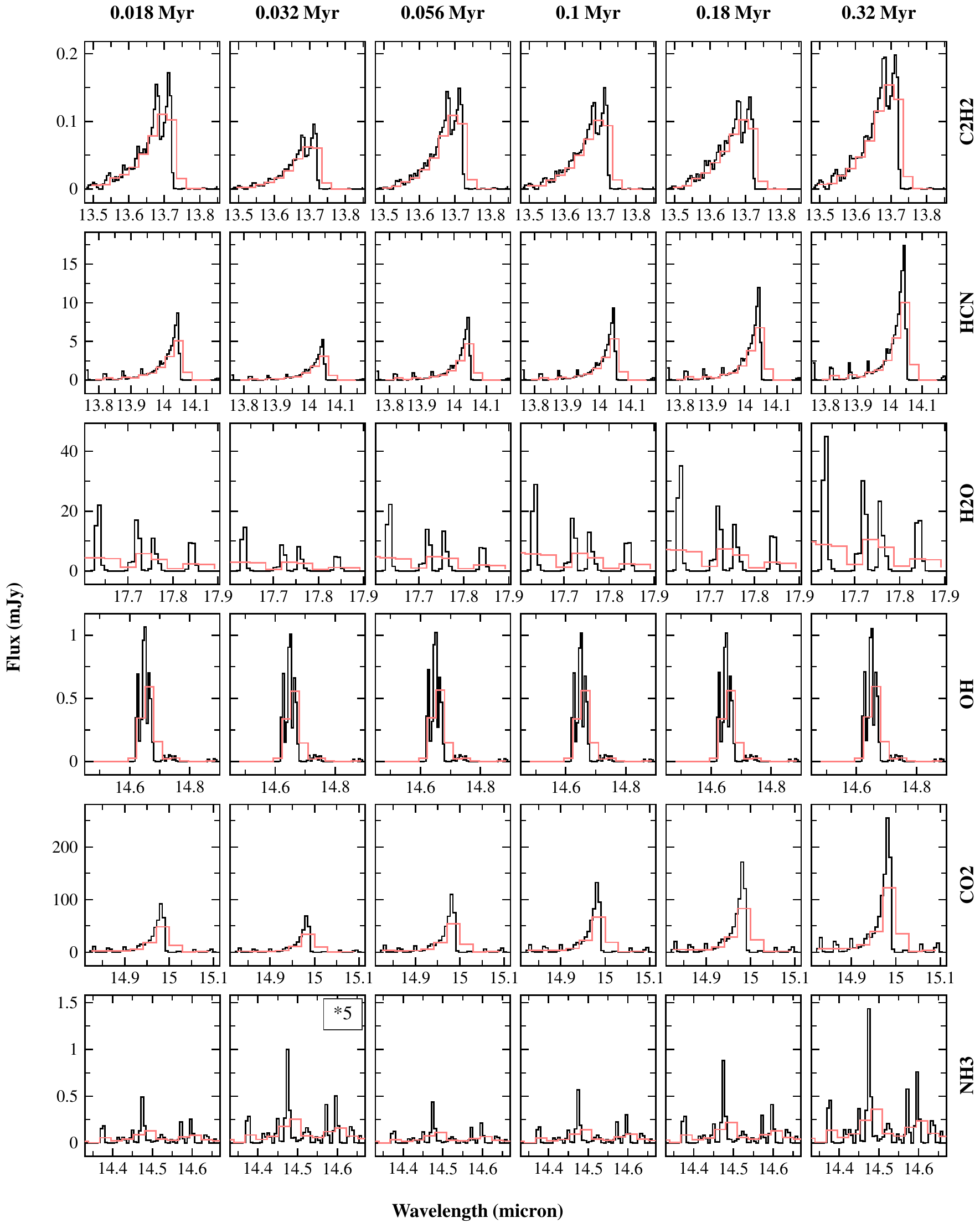}
			\caption{\flits{} spectra of the disk models up to $0.32~\myr$ old, convolved to a spectral resolution $R=2\,800$ (black lines) and $R=600$ (red lines). Where indicated, the spectra have been multiplied by $5$, $10$, or $50$ in order to improve visibility. This disk is at an inclination of $45^\circ$. The reason why the spectral lines become weaker at $0.032~\myr$, only to become stronger again at $0.056~\myr$, is likely due to the dust evolution: \cref{fig:radial_properties_combined} shows that radii smaller than $1~\au$, the gas-to-dust ratio drops between the $0.018~\myr$ and $0.032~\myr$ models, before increasing again at the $0.056~\myr$ mark. Because the onset of radial drift suddenly decreases the gas-to-dust ratio, we see a corresponding drop in line fluxes.}
			\label{fig:flitsspec1}
		\end{figure*}

\begin{figure*}
			\centering
			\includegraphics{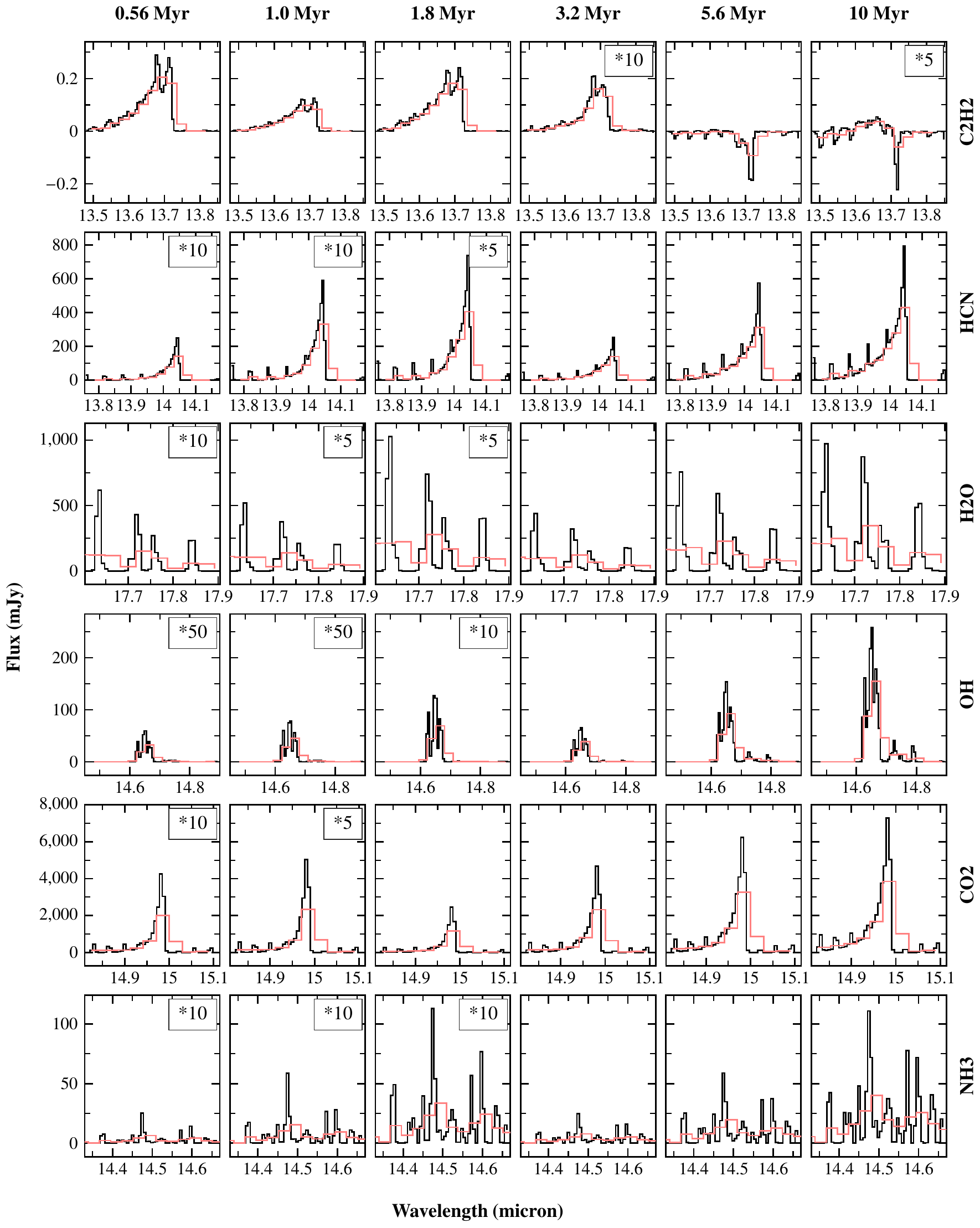}
			\caption{\flits{} spectra of the disk models from $0.56~\myr$ old, convolved to a spectral resolution $R=2\,800$ (black lines) and $R=600$ (red lines). Where indicated, the spectra have been multiplied by $5$, $10$, or $50$ in order to improve visibility. This disk is at an inclination of $45^\circ$. The $y$-axis limits are different to those in \cref{fig:flitsspec1}}
			\label{fig:flitsspec2}
		\end{figure*}

		  \Cref{fig:perspecies_combined} shows that where we compute the spectra of our models using \flits{}, we expect the fluxes of every species except \cem{C2H2} to increase significantly as the dust in the disk evolves.  This is illustrated by \cref{fig:flitsspec1,fig:flitsspec2}, which show that between the youngest and oldest disks models, the peak flux densities for every species except \cem{C2H2} increase by factors ranging from 30 for \cem{H2O} to 200 for \cem{OH}. The escape probability line fluxes in \cref{fig:perspecies_combined} increase by similar amounts, showing that measuring the escape probability line flux of these single lines forms a reasonable proxy for measuring trends in the overall strength of the line-emitting region.

		For comparison to infrared observatories, the median $1~\sigma$ mid-infrared line sensitivity of \spitzer{} is about \mbox{$8.5 \times 10^{-19}~\mathrm{W~m}^{-2}$}, thus many individual spectral lines are far too faint for \spitzer{} to detect.\footnote{\url{irsa.ipac.caltech.edu/data/SPITZER/docs/irs/}.}
		However, the ro-vibrational bandheads of species such as \cem{CO2} contain many lines that at a low spectral resolution are blended together, which is why we can see these species with \spitzer{}. In some of the more evolved disks, even individual \cem{OH}, \cem{o-H2O}, \cem{NH3}, and \cem{HCN} lines are significantly above this sensitivity limit (see \cref{fig:perspecies_combined}).
Following the flux densities of the convolved spectra presented in \cref{fig:flitsspec1,fig:flitsspec2},  and assuming a $5~\sigma$ \spitzer{} sensitivity of $5~\mjy$ in $512~\mathrm{s}$,  we expect that only \cem{C2H2} would be undetectable at every age. \cem{OH}, \cem{H2O}, and \cem{NH3} are only detectable in more evolved disks, while \cem{HCN} and \cem{CO2} should be detectable in every model. At $R=600$, the complexes of water lines can easily become blurred at \spitzer{}'s low spectral resolution, making it difficult to determine the continuum flux (in comparison to $R=2800$ spectra).  The $10~\sigma$ continuum sensitivity of \jwst{}'s MIRI instrument is around $0.1~\mjy$ at $15~\micron$, with an exposure time of $10^4~\mathrm{s}$ \citep{Glasse:2015cg}. Thus with \jwst{}, we can expect that in every model, all species except \cem{C2H2} would be detectable with exposure times much less than $10^4~\mathrm{s}$.

\cem{C2H2} even goes into absorption at $5.6~\myr$ and $10~\myr$. The line emission features in disk models up to $1.8~\myr$ old appear to be dominated by emission from near the inner wall, while in the older disk models the line-emitting region extends out to $0.6~\au$ and is close to the mid-plane. The absorption lines in the $5.6~\myr$ and $10~\myr$ old models are likely attributable to the fact that the \cem{C2H2} emission comes from close to the mid-plane. As described in {Greenwood et al. (submitted)}, there exists an inversion in the gas temperature around the $\av=1$ line, such that there may exist clouds of colder gas above the warmer gas in the disk. Absorption lines may result if these clouds of colder gas exist between our line of sight and the line-emitting region of the species.

\cem{CO2} is the brightest species by a significant margin, suggesting that our younger models may be representative of the sub-class discovered by \cite{Pontoppidan:2010gw}, where in six disks only \cem{CO2} was unambiguously detected. This sub-class may consist of disks where the dust is significantly settled in comparison to the gas, similar to the structure produced in our models by the self-similar dust settling. In this case, we would expect \cem{CO2} to be the brightest species in the mid-infrared, and thus other species may remain undetected. The reason that \cem{CO2} fluxes can be so high is that it is a robust species which is able to survive in relatively optically thin regions because it has a small cross-section for photodissociation \citep{vanDishoeck:2006hc}.  As the dust structure evolves and becomes more optically thin, the line-emitting area of \cem{CO2} also grows: \cref{fig:perspecies_combined} shows that \cem{CO2} can have a line-emitting area close to $1000~\au^2$ (the \cem{CO2} emission extends to about $15~\au$ at $1~\myr$).


		\section{Discussion}

		As the dust evolves, grains are depleted from the disk and the mid-IR molecular line emission grows substantially for every species except \cem{C2H2}. This happens because dust particles (particularly small micron-sized grains) are the main carriers of opacity in the disk, so reduced continuum optical depths lead directly to a larger line-emitting area and increased molecular line fluxes.

		The lack of \cem{C2H2} emission is because this species exists closer towards the mid-plane than the other molecules. Thus, it is more affected by both the optical depth of the dust in the disk and the gas temperature inversions around $\av=1$. The T~Tauri spectra observed by \citet{Carr:2011hl} have peak \cem{C2H2} flux densities of up to about $20~\mjy$, and \cem{C2H2}/\cem{HCN} line flux ratios between about $0.08$ and $0.8$.  As \cref{fig:perspecies_combined,fig:flitsspec1,fig:flitsspec2} show, when we compare the peak flux densities of the \flits{} spectra of our strongly-evolved models, we see much lower ratios: although the ratio between the spectral flux densities of \cem{C2H2} and \cem{HCN} is $\sim 0.02$ at $0.18~\myr$, it decreases with age to as low as $5\times 10^{-5}$ at $10~\myr$ (with the caveat that for the oldest models, \cem{C2H2} is in weak absorption).  The fact that \cem{C2H2} behaves so differently to other species suggests either that our chemical network misses some key \cem{C2H2} formation pathways, or that the dust structure of our models is not representative of disks with \cem{C2H2} detections.
		Although it is possible that uncertainties in the chemical network and reaction rates are leading to significant inaccuracies in the distribution of \cem{C2H2} calculated by \prodimo{}, our \cem{C2H2}-deficient models may also be a natural result of the dust structure.   \citet{Walsh:2015jr} find that \cem{C2H2} is significantly less abundant in Herbig Ae disks due to increased levels of photodissociation, which also increases when the population of small dust grains is reduced.   This suggests that the low \cem{C2H2} line fluxes are a result of an evolved dust structure, but it does not adequately explain the fact that \cem{C2H2} is all but absent in even our youngest $0.018~\myr$ model. 
		
		Another notable result is that \cem{CO2} is much brighter than all other species in our models, suggesting that our older models may be representative of the disks found by \citet{Pontoppidan:2010gw} where only \cem{CO2} was detected: if these sources have an evolved and settled dust structure, our results show that we would expect \cem{CO2} to be much brighter than other species. The sources where \cem{CO2} was not the only detected species may have a less evolved and settled dust structure, giving lower \cem{CO2} line fluxes and more equal ratios between the line fluxes of different species.
		In all of our models, the \cem{CO2} spectrum has a brighter flux density than all other species by at least a factor of ten.
		The explanation for the relative brightness of \cem{CO2} likely lies in the dust and temperature structure of the models.
		
		While \citet{Pontoppidan:2010gw} note a strong correlation between spectral type and line detections, there also appears to be a dichotomy in their results between B, A, and F type objects and G, K, and M spectral types. For spectral types earlier than G, there is a dramatic and as-yet unexplained drop in the detection rates of all species except \cem{CO}.
		This may appear to be in conflict with  \citet{Mulders:2012ew}, who suggest that  late-type (brown dwarf and T~Tauri) disks have flatter dust disks and relatively thicker gas disks than earlier (Herbig) disks do.\footnote{One caveat to their analysis is that the best-fit brown dwarf, T~Tauri, and Herbig disk models have identical outer radii. Their brown dwarf model has a very low surface density and is strongly flared, while the outer radius of $400~\au$ is not representative of a typical brown dwarf disk \citep{Bate:2003cv}.} In this paper we suggest that a highly-evolved and strongly-settled dust structure may lead to \cem{CO2} emission being much brighter than other species. However, if younger disks have less-evolved dust (thus more small dust grains) and a less-settled dust structure, and because our models show that \cem{CO2} line fluxes increase with age more than other species, we suggest that the flux ratios between \cem{CO2} and the other species that we have modelled will be lower in early-type disks.
		
		The work by \citet{Mulders:2012ew} is based upon SEDs, and they state that ``regions with the same temperature have a self-similar vertical structure independent of stellar mass''. It is difficult to distinguish changes in the SEDs of our dust evolution models at wavelengths below $30~\micron$; thus we are probing different parameter spaces. The lack of detections around early-type stars could be because the stronger UV radiation of early-type stars is enough to prevent species that are vulnerable to photodissociation from forming in sufficient quantities.  Such species include \cem{C2H2}, \cem{H2O}, \cem{NH3}, and \cem{HCN}, which all have photodissociation cross sections $>10^{-17}~\mathrm{cm}^{-2}$ \citep{vanDishoeck:2006hc}.  \citet{Walsh:2015jr} find that weaker UV fields do indeed allow for more molecule-rich disk atmospheres. Thermochemical models by \citet{Antonellini:2015ir} have found that strong UV radiation does not weaken water line fluxes and that likely explanation for the lack of water detections around Herbig stars is noisy spectra combined with high continuum flux levels. Future modelling and observational work is required in order thoroughly to understand the interplay between the disk dust structure, the radiation field, and the mid-infrared molecular lines. What appears most clear is that as the main opacity carrier in the disk, the evolution and settling of dust can have a substantial impact.
		
		\section{Conclusion}
		
We have coupled the dust evolution code \twopoppy{} with the radiative transfer code \mcmax{}, the thermochemical disk modelling code \prodimo{} and the line-tracing code \flits{}. We have produced a series of models that simulate the evolution of dust over time in a $2$D thermochemical disk model, and have created infrared spectra of these models including \cem{C2H2}, \cem{CO2}, \cem{HCN}, \cem{NH3}, \cem{OH}, and \cem{H2O} mid-infrared line emission.
		
This paper shows that dust evolution has a very clear and straightforward effect on the mid-infrared line fluxes, where the evolving and diminishing dust population in the surface acts to decrease continuum optical depths and thus increase line fluxes by up to a factor of $100$. Even in the least-evolved disk models, the mid-infrared lines prefer to emit from a gas-to-dust ratio of at least $1000{:}1$. This provides a physical explanation for the need in previous literature to use such ratios in order to produce enough line flux to match models with observations.

This is a double-edged sword: although the dust structure is a significant degeneracy when fitting models to observations, if we can observe the dust directly (for example, by observing the $\tau=1$ surface with scattered-light imaging), the dust structure also becomes a powerful diagnostic in understanding the mid-infrared emission of T~Tauri disks.  Upcoming observatories such as \jwst{} and \eelt{} will significantly improve our ability to observe these disks, and combined with more sophisticated models, we can build a two- or three-dimensional understanding of dust structure in the inner disk and the mid-infrared spectral lines.

		\section{Acknowledgements}
		
		We would like to thank Lucia Klarmann, Gabriela Muro Arena, Paola Pinilla, Til Birnstiel, and Michiel Min for their discussions and comments on the manuscript. We also thank  the Center for Information Technology of the University of Groningen for their support and for providing access to the Peregrine high performance computing cluster.

		\bibliographystyle{aa}
		\bibliography{fullbib_local}

	\end{document}